\documentclass{article}
\usepackage{amsmath,amsfonts,amssymb,amsxtra}
\usepackage{graphicx}

\newcommand{\InvA}{\AA$^{-1}$}
\newcommand{\grad}{\ensuremath{^\circ}}

\newcommand{\lco}{La$_{2}$CoO$_{4}$}
\newcommand{\lsco}{La$_{2-x}$Sr$_{x}$CoO$_{4}$}

\newcommand{\prca}{Pr$_{2-x}$Ca$_{x}$CoO$_{4}$}

\newcommand{\lscoDD}{La$_{5/3}$Sr$_{1/3}$CoO$_{4}$}
\newcommand{\lscoIV}{La$_{1.6}$Sr$_{0.4}$CoO$_{4}$}
\newcommand{\lscoV}{La$_{1.5}$Sr$_{0.5}$CoO$_{4}$}

\newcommand{\htsc}{La$_{2-x}$Sr$_{x}$CuO$_{4}$}
\newcommand{\lsno}{La$_{2-x}$Sr$_{x}$NiO$_{4+\delta}$}
\newcommand{\lsnoIV}{La$_{1.6}$Sr$_{0.4}$NiO$_{4}$}

\newcommand{\coii}{Co$^{2+}$}
\newcommand{\coiii}{Co$^{3+}$}

\newcommand{\lscoXII}{La$_{1.88}$Sr$_{0.12}$CuO$_{4}$}

\title{Hour-glass magnetic spectrum in a stripe-less insulating transition metal oxide}

\author{Y. Drees$^{1}$, D. Lamago$^2$, A. Piovano$^3$ \& A.~C.~Komarek$^1$}

\begin{document}

\maketitle

\begin{enumerate}
 \item Max-Planck-Institute for Chemical Physics of Solids, N\"{o}thnitzer Str. 49, D-01187 Dresden, Germany
 \item Laboratoire L\'{e}on Brillouin, CEA/CNRS,F-91191 Gif-sur Yvette Cedex, France
 \item Institut Laue-Langevin (ILL), 6 Rue Jules Horowitz, F-38043 Grenoble, France
\end{enumerate}

\begin{centering}
\textbf{
An hour-glass shaped magnetic excitation spectrum appears to be an universal characteristic of the high-temperature superconducting cuprates.
Fluctuating charge stripes or alternative band structure approaches are able to explain the origin of these spectra.
Recently, an hour-glass spectrum has been observed in an insulating cobaltate,
thus, favouring the charge stripe scenario.
Here we show that neither charge stripes nor band structure effects are responsible for the hour-glass dispersion in a cobaltate within the checkerboard charge ordered regime of \lsco.
The search for charge stripe ordering reflections yields no evidence for charge stripes in \lscoIV\ which is supported by our phonon studies.
With the observation of an hour-glass-shaped excitation spectrum in this stripe-less insulating cobaltate, we provide experimental evidence that the hour-glass spectrum is neither necessarily connected to charge stripes nor to band structure effects, but instead, probably intimately coupled to frustration and arising chiral or non-collinear magnetic correlations.}
\end{centering}

\par \vspace{0.5cm}
\par
Charge stripes that have been initially predicted as a combined charge- and spin-density wave phenomenon \cite{stripesA,stripesB,stripesC}
have been first observed experimentally in Nd-codoped \htsc\ Ref.~\cite{stripesLNSCO} and in the isostructural nickelates \lsno\ (LSNO) Ref.~\cite{stripesLSNO} with a quasi-two-dimensional layered structure.
These charge stripes are characterized by hole-rich regions running in vertical \cite{stripesLNSCO} or diagonal \cite{stripesLSNO} direction within the $M$O$_2$ planes ($M$~=~Cu/Ni), thus, separating the remaining charge depleted regions where the antiferromagnetic (AFM) structure of the undoped parent compound recovers.
Since charge stripes act as antiphase domain walls, the magnetic structure appears modulated.
The role of this charge stripe instability for the high-temperature superconducting (HTSC) cuprates remains a matter of debate
and it is puzzling that both phenomena can coexist though charge stripes tend to suppress superconductivity \cite{stripesVSsc}.
Another isostructural system where stripe ordering has been reported is \lsco\ Ref.~\cite{cwik,hourglass}. Surprisingly, also an hour-glass magnetic spectrum has been observed in \lscoDD\ Ref.~\cite{hourglass}. This hour-glass spectrum resembles the famous excitations in the cuprates \cite{hourglassLSCO,hourglassLSCOb,hourglassLSCOc,hourglassLSCOd,hourglassLSCOe,hourglassLSCOf,hourglassLSCOg,hourglassLSCOh,hourglassLSCOi,pseudo} which were found to be a universal feature of these HTSC materials and which has stimulated enormous efforts for explaining these spectra \cite{hourglassExplA,hourglassExplB,hourglassExplC,hourglassExplD}. Very recently, theoretical simulations of the magnetic excitation spectra based on a disordered charge-stripe model have been also reported for \lscoDD\ Ref.~\cite{andrade}.
These layered cobalt oxides are isostructural to the prototypical cuprate materials \htsc. The quasi-two dimensional crystal structure arises from a stacking of CoO$_2$ layers that are separated by (La,Sr)O rocksalt layers acting as a charge reservoir. The substitution of trivalent La by divalent Sr introduces holes into the CoO$_2$ layers. In analogy with the isostructural nickelates (LSNO) these holes were proposed to segregate into diagonal charge stripes \cite{cwik,hourglass}. However, whereas the incommensurate magnetic peaks can be clearly observed in \lsco\ Ref.~\cite{cwik} the corresponding unambiguous observation of sharp and well separated charge-stripe ordering (CSO) superstructure reflections is missing. Hence, also a glassy charge ordered state with different commensurate superlattice fragments could be responsible for the observed broad signals in \prca\ Ref.~\cite{highdoped} or even simply disordered magnetic contributions spread around half-integer positions in reciprocal space for the reported features in \lsco\ Ref.~\cite{cwik}.

\par Here, we have studied the magnetism, magnetic excitations, charge ordering and electron phonon coupling in a cobaltate compound close to half-doping, i.e. in \lscoIV\ and were able to observe the appearance of an hour-glass dispersion in this stripe-less cobaltate within the checkerboard-charge ordered regime of \lsco.

\section{Results}
\subsection{Spin and charge order}
\par In contrast to other studied cobaltates within the incommensurate magnetic regime our sample exhibits comparably sharp incommensurate magnetic peaks and, thus, much less disorder than in previously studied compounds (see Fig.~1). Note, that it is shown in Ref.~\cite{andrade} that the sharpness of the CSO signal should be roughly equal to the sharpness of the incommensurate magnetic peaks in a CSO scenario for \lsco\ Ref.~\cite{andrade}. Hence, our \lscoIV\ material seems to be an ideal system to study any possible relationship between incommensurate magnetic peaks and charge stripes in these layered cobaltates.  The results of our elastic neutron scattering experiments on \lscoIV\ are shown in Fig.~1. Finally, we were not able to detect any signal indicative for CSO in \lscoIV, see Fig.~1~(c). Instead, only one broad peak at half-integer positions appears that is indicative for a (disordered) checkerboard charge ordered (CBCO) state persisting up to high temperatures very similar to the optimum half-doped compound \lscoV\ (dashed line). Thus, we managed to find a \lsco\ sample within the CBCO ordered regime which exhibits clearly incommensurate magnetism.

\subsection{Magnetic excitation spectra}
\par Next, we studied the magnetic excitations of this compound.
In Fig.~2~(a-f) the magnetic excitation spectra of \lscoIV\ are shown and compared with those in \lscoV.
In contrast to the half-doped reference material (the dashed lines in Fig.~2~(d,f) were derived from spin wave calculations using the McPhase program code \cite{mcphas} closely following the analysis in Ref.~\cite{half}), all basic features of an hour-glass magnetic spectrum can be observed in \lscoIV\ - an inwards-dispersion of low-energy branches towards the planar AFM wavevector, a suppression of the outwards-dispersing branches, a resonance-like merging around $\sim 20$~meV with increased intensity and an outwards dispersion of the high energy branches measured at positions rotated by 45\grad\ compared to the elastic magnetic satellites.
Additional constant-Energy maps for \lscoIV\ are shown in Fig.~2~(g-o). Remarkably, the high energy magnetic excitations in \lscoIV\ resemble on the isotropic high energy excitations recently reported for La$_{1.96}$Sr$_{0.04}$CuO$_4$ Ref.~\cite{MagLSCOisotropic}. The constant-$Q$ scan at ($3/2$~$1/2$~0) clearly shows the resonance-like increase of intensity at the magnon merging point, see Fig.~2~(p). Furthermore, we have studied the low-energy magnetic excitations at a cold triple-axis spectrometer. In Fig.~2~(q) the energy-scan at an incommensurate magnetic peak position is shown which indicates the possible presence of a gap in the spin excitation spectrum that amounts to about $\sim$1.8~meV. The incomplete suppression of inelastic magnetic intensities below this possible gap could be explained by an intrinsic peak width of the magnetic modes arising from short-lived damped magnetic excitations (that are probably induced by the high degree of disorder by the random doping of electrons into the CBCO matrix). The vicinity of our 40\% Sr-doped sample to the half-doped parent compound might favour a similiar origin of the gap as in \lscoV, i.e. a single ion or small exchange anisotropy within the $ab$ plane \cite{half}. Also the existence of a gap in these hour-glass shaped excitations in \lscoIV\ resembles on the observations in the isostructural HTSC cuprates (e.g. in \lscoXII\ Ref.~\cite{LSCOgap}) even if their origin might be different.

\subsection{Electron-phonon coupling}
\par We have also searched for signatures of charge ordering in the  Co-O bond stretching phonon dispersion of \lsco. In Fig.~3~(a) two polarization patterns of  high-frequency $\Sigma_1$ phonon modes are shown. In the left figure it is shown schematically that the presence of diagonal charge stripes (indicated by the dashed lines) would induce a coupling of the Co-O bond stretching phonon mode to stripes with the same propagation vector. That charge stripes in single layer perovskite oxide materials are able to induce such an electron phonon coupling and even giant electron phonon anomalies was already demonstrated for high-energy Cu-O bond stretching phonon modes in the isostructural cuprates \cite{LSCOphonon}. On the other hand, for CBCO one would expect bond stretching phonon anomalies at the zone boundary as is shown schematically in the right part of Fig.~3~(a).
Here, we have measured the topmost Co-O bond stretching phonon dispersion of \lscoIV\ and several \lsco\ reference samples ($x$~$=$~$0$, $1/3$, $0.5$), see Fig.~3~(b,c,e).
Compared to the undoped \coii\ compound \lco\ the \lsco-materials within the incommensurate magnetic regime all exhibit a strong pronounced softening of the phonon dispersion towards the zone boundary i.e. at ($1/2$~$1/2$~0) and not at propagation vectors corresponding to the expected position for CSO at ($\zeta$~$\zeta$~0) with $\zeta$~$=$~$2\cdot\varepsilon$~$\sim$~$x$ (indicated by the black arrows in Fig.~3~(c,d)).
In contrast to our observations in \lscoIV\ the corresponding Ni-O bond stretching phonon mode of a 40\%~Sr-doped LSNO reference material with robust diagonal CSO exhibits a completely different and almost complementary dispersion, see Fig.~3~(d). The high-frequency Ni-O bond stretching mode starts at very similar energies for $\zeta$~$=$~$0$ but quickly softens around propagation vectors corresponding to CSO propagation vectors and even exhibits a final upturn towards the zone boundary. Note, that we also observed phonon softening in other LSNO samples (with $x$$=$$0.2$) where the phonon softening appears at phonon propagation vectors that exactly correspond to the propagation vectors of charge stripe ordering of that particular LSNO sample \cite{dissACK}.
Hence, an opposing behaviour of bond stretching phonon modes in cobaltates and in charge stripe ordered nickelates can be observed. In the cobaltates, the anomalous phonon softening at the zone boundary is consistent with a robust CBCO persisting also below half-doping rather than with static (or dynamic) charge stripe phases and supports our elastic studies.

\section{Discussion}
\par The absence of stripes in \lscoIV\ and our observation of an hour-glass-shaped excitation spectrum in this compound (even with indications for a gap) sheds another light on the recent argumentation that the absence of any gap in the hour-glass-shaped excitations of the cobaltates (\lscoDD) with static stripe order and the appearance of a gap in the excitations of the HTSC cuprates points to a ``collective quantum melting of stripe-like electronic order'' in the cuprates \cite{zaanen}.
Also the study of larger parts of the phase diagram of \lsco\ motivates a different mechanism for the emergence of incommensurate magnetism in \lsco\ ($x$~$\lesssim$~$1/2$), see Fig.~4.
Apparently, the optimum doped CBCO ordered compound \lscoV\ exhibits the sharpest magnetic peaks which broaden with increasing distance of hole-doping away from half-doping. Similar observations have been reported in Ref.~\cite{cwik}. With decreasing Sr-doping the incommensurate magnetic regime finally breaks down slightly below $1/3$-doping.
Compared to a CSO scenario the continuously increasing peak width with decreasing Sr-concentration below half-doping in the cobaltates
is more indicative for a rising amount of frustration in an alternative chiral or non-collinear magnetic scenario: starting from the optimum half-doped composition with sharpest magnetic peaks the doping of additional electrons (\coii-ions) into the ideal CBCO ordered AFM matrix induces frustration. Thus, the substitution of \coii-ions for non-magnetic \coiii-ions alters the AFM \coii$^{\uparrow}$-\coiii-\coii$^{\downarrow}$-\coiii-\coii$^{\uparrow}$ exchange paths resulting in frustrated spin arrangements \coii$^{\uparrow}$-\coii$^{\downarrow}$-\coii$^{?}$-\coiii-\coii$^{\uparrow}$, see Fig.~4~(d). In order to relieve the high amount of frustration that is generated by the introduction of additional large nearest neighbour exchange couplings $J$~$>>$~$J'$ the system can be expected to turn into a chiral or non-collinear magnetic state. This scenario is not only in agreement with our observation of incommensurate magnetism together with the absence of charge stripes but it is also able to explain naturally the successively increasing peak broadening away from half-doping, see Fig.~4~(b).
Also the kind of incommensurate magnetism naturally follows from our frustration scenario. As shown in Fig.~4~(d) each doped electron into the undistorted AFM matrix of the ideal checkerboard charge ordered half-doped system induces additional strong AFM nn-exchange interactions. Therefore, the nnn-interactions become ferromagnetic (FM) in total. Without electron doping, these exchange interactions all had been AFM before. In order to release frustration, the magnetic moments start twisting/spiralling (a spin-density wave appears unlikely in this strongly localized system). The more electrons are doped within a certain given distance in $a$-direction (in $b$-direction), the more centers of frustration and the more spiralling occurs within this certain distance in order to release frustration. This machanism would be consistent with the increasing incommensurability away from half-doping. Furthermore, it is also able to explain the direction of the incommensurate magnetic satellites: Since these spiraling-effects appear concomitantly in $x$- and in $y$- direction (in-plane), the diagonal direction of the incommensurate satellites arises from the presence of two independent AFM sublattices in the ideal optimum half-doped compound (see \cite{half}) which are affected both together at the same time by the additional doping of one electron, see Fig.~4~(d). Since there will be the same total twist of spins in $a$-direction and in $b$-direction after a certain distance the total modulation i.e. the propagation appears to be in $[$$\pm$$1$~$\pm$$1$~$0$$]$-direction. Hence, in this particular cobaltate system with two independent AFM sublattices frustration is able to explain the direction of the incommensurate magnetic satellites as well as the increasing amount of spiralling with increasing electron doping away from half-doping and, thus, the continuously increasing incommensurability.
In contrast to our frustration scenario a charge stripe scenario will probably have problems in explaining the increasing peak broadening of the incommensurate magnetic peaks with decreasing hole-doping from $x=1/2$ towards $x=1/3$ since the most stable diagonal stripe phase with sharpest magnetic reflections can be expected for $x=1/3$ as in the nickelate case. But exactly the opposite can be observed in these layered cobaltates. Note, that the amount of disorder in the simulations of \cite{andrade} was introduced fully artificially.
A chiral magnetic state has been also proposed for the insulating spin glass phase of \htsc\ \cite{LSCO,LSCOb,hasselmann} exhibiting also diagonal magnetic satellites and magnetic excitations with the characteristics of an hour-glass spectrum \cite{hourglassLSCOb}. In this chiral model of the magnetic ground state of \htsc\ frustration arises from doping of charges into the AFM matrix of an AFM parent compound (La$_2$CuO$_4$) Ref.~\cite{hasselmann,LSCO,LSCOb}. Note, that also in bulk \htsc\ no CSO peaks have been found \cite{ourLSCO}, and, that in both isostructural systems ($M$~$=$~Cu or Co respectively) frustration might arise from the doping of additional charges (holes or electrons respectively) to interstitial sites, thus, effectively introducing additional ferromagnetic exchange interactions into the unaltered AFM host structure (see Ref.\cite{hasselmann} or Fig.~4~(d) respectively). Hence, frustration might not only play an important role for the triangular lattice \cite{nacoo2} but also for the hole-doped square lattice systems as well as for the emergence of the hour-glass dispersion in these systems.

\par Concluding, we studied the magnetic excitations and electron phonon coupling of a single layer perovskite cobaltate within the checkerboard charge ordered regime. In this material with comparably sharp incommensurate magnetic peaks we observe the appearance of a magnetic excitation spectrum with all basic features of an hour-glass spectrum in close vicinity to half-doping. The absence of any charge stripe superstructure reflections in this material indicates that the hour-glass dispersion might emerge concomitantly with the onset of frustration and chiral or non-collinear magnetic ordering in \lsco\ ($x$~$\lesssim$~$1/2$) rather than with the onset of charge stripe ordering. We also studied the impact of electron-phonon coupling in these materials and were able to observe large Co-O bond stretching phonon anomalies at the zone-boundary which are consistent with our picture that robust checkerboard charge ordering is still prevalent below half-doping in \lsco. Neither the existence of static nor of dynamic charge stripe phases can be supported by these high-frequency phonon studies. Our observation of an hour-glass-shaped excitation spectrum in a stripe-less insulating transition metal oxide goes one step further than the recent observation of an hour-glass spectrum in an insulating cobalt oxide \cite{hourglass} since it clearly shows that besides Fermi-surface effects also no charge stripes are needed for the appearance of the hour-glass spectrum. Therefore, frustration might be the unifying property of all these systems that exhibit an hour-glass-shaped magnetic excitation spectrum and either band structure effects or (short ranged) charge stripe correlations are just additional side effects which are not directly connected to the appearance of this peculiar excitation spectrum.

\section{Methods}
\subsection{Synthesis}
Single crystals of \lsco\ and LSNO materials have been grown by the floating zone technique following a route as described elsewhere \cite{growthLSCoO,LSNOcryst}. But in order to determin the oxygen content more carefully we have grown all the crystals in Ar atmosphere. Only one of our single crystals, i.e. the nominally 1/3 Sr-doped compound, has been grown in an oxygen enriched Ar-atmosphere  in order to obtain one single crystal that was grown under more oxidizing conditions like described for \lscoDD\ in \cite{hourglass}.
The high single crystal quality has been ascertained as described in the Supplementary Note~1.
\subsection{Neutron scattering}
For most inelastic neutron measurements two or three large single crystals have been co-aligned with the crystallographic $a_{\rm tet}$- and $b_{\rm tet}$-axes in the scattering plane.
For measuring the magnetic excitation spectra, inelastic neutron scattering experiments have been performed at the 2T.1
triple-axis spectrometers (TAS) at the LLB in Saclay, France.
A vertically focussing configuration of the PG-$002$ analyzer and a flat configuration of the PG-$002$ monochromator has been chosen and higher order contributions have been suppressed by two PG-filters. All inelastic scans have been performed in the constant-$k_f$-operation mode with $k_{\rm f}$~=~2.662~\InvA\  in most cases and $k_{\rm f}$~=~4.1~\InvA\ for the highest measured energies solely.
For measuring the high-frequency $\Sigma_1$ phonon dispersion and the constant energy maps, inelastic neutron scattering experiments have been performed at the IN8 thermal TAS at the ILL in Grenoble, France.  A Cu$200$ double-focusing monochromator and a PG-$002$ double-focusing analyzer have been chosen for measuring these high-energy phonon modes in constant-$k_{\rm f}$-operation mode with $k_{\rm f}$~=~2.662~\InvA. Higher order contamination was suppressed by two PG filters. For measuring the constant energy maps in Fig.~2 a Si-111 monochromator has been used in order to suppress $\lambda$/2 contaminations. Powder lines originating from the cryostat and the sample holder are apparent in these measurements. In order to apply a background-correction we have also performed measurements with empty sample holder (measurements without sample) for the most important constant-E maps: first, at the neck of the hour-glass, i.e. at 20 meV, and, second, at a distinctly higher energy of 27 meV. Thus, we were able to subtract the background from our sample signal in order to confirm the shape of our observed magnetic signal. These two maps are shown in Fig.~2~(n,o). Note that the spurious peaks could still not be corrected this way since they might originate from monochromator phonons that were scattered elastically by the sample etc.
The magnon merging point as well as the outwards dispersion can be clearly seen in these kind of (HK0)-maps at 20~meV and 27~meV respectively.
Additional inelastic neutron scattering experiments of the low energy magnetic excitations have been performed at the cold neutron TAS 4F.2 at the LLB in Saclay, France ($k_{\rm f}$~$\sim$~1.55~\InvA; Be-filter at 77~K).
Elastic neutron scattering measurements have been performed at the 3T.1 diffractometer and 1T.1 TAS at the LLB in Saclay, France and at the IN8 TAS spectrometer at the ILL in Grenoble, France ($k_i$~$\sim$~2.662~\InvA; two PG filters have been used for suppression of higher order contamination).
%All error bars shown in Figs.~1-4 are standard deviations of Gaussian peak fits to the data, and, if neutron scattering intensities are shown, were calculated by the square root of the measured intensity.

\begin{itemize}
 \item \textbf{Acknowledgements} We thank Y. Sidis, D. I. Khomskii, E. Andrade, M. Rotter, P. Thalmeier, O. Stockert and L. H. Tjeng for helpful discussions. We thank H. Borrmann, Y. Prots and S. H\"{u}ckmann for X-ray diffraction measurements.
     We thank S.~Subakti for titration measurements and we thank A.~Keil for help in cutting crystals. We thank the team of U.~Burkhardt for EDX measurements and the team of G.~Auffermann for ICP measurements.
 \item \textbf{Contributions} A.~C.~K. planned all experiments. A.~C.~K. synthesized all studied materials. Y.~D. and D.~L. performed the elastic and inelastic neutron scattering experiments at the 3T.1 diffractometer and at the 2T and 1T spectrometers. Y.~D., A.~P. and A.~C.~K. performed the elastic and inelastic neutron scattering experiments at the IN8 spectrometer.
     A.~C.~K. wrote the manuscript.
 \item \textbf{Competing Interests} The authors declare that they have no
competing financial interests.
 \item \textbf{Correspondence} Correspondence should be addressed to A.~C.~K.~(email: Komarek@cpfs.mpg.de).
\end{itemize}

\begin{figure}[ht]
\begin{center}
\includegraphics[width=0.8\textwidth]{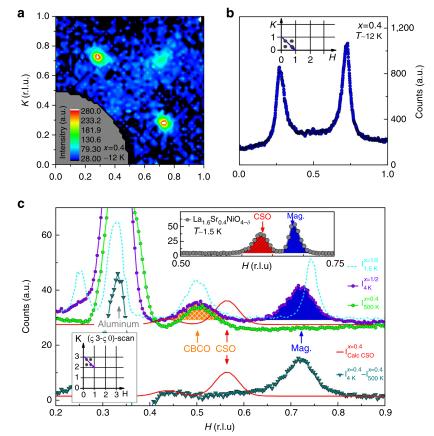}
\end{center}
\caption{\textbf{Elastic neutron diffraction study of charge and magnetic correlations} (a) Neutron scattering intensities in the $HK0$-plane for \lscoIV\ at 12~K; (200~K background subtracted). (b) Magnetic peaks measured in scans across (1/2~1/2~0). (c) Diagonal ($H$~$3$$-$$H$~$0$)-scans for \lscoIV\ measured at 4~K and at 500~K. Also the difference of 4~K and 500~K intensities is shown together with the expected CSO intensity (red line) derived from a CSO-ordered LSNO reference sample (inset) with the same Sr-doping ($I_{calc\ CSO}^{x=0.4}=\alpha\cdot I_{magn.}^{x=0.4}\cdot I_{CSO}^{LSNO}/I_{magn.}^{LSNO}$; $\alpha=(m_{Ni}/m_{Co})^2\cdot (u_{Co}/u_{Ni})^2 \sim  1$). The clear absence of any CSO signal and the broad signal at half-integer positions indicates that our \lscoIV\ sample is still within the CBCO regime. Dashed cyan line: the charge ordering signal of the optimum doped CBCO ordered sample \lscoV. Intensity error bars are statistical error bars calculated by the square root of intensity.}
\label{fig5}
\end{figure}

\begin{figure}[ht]

\begin{center}
\includegraphics[width=0.8\textwidth]{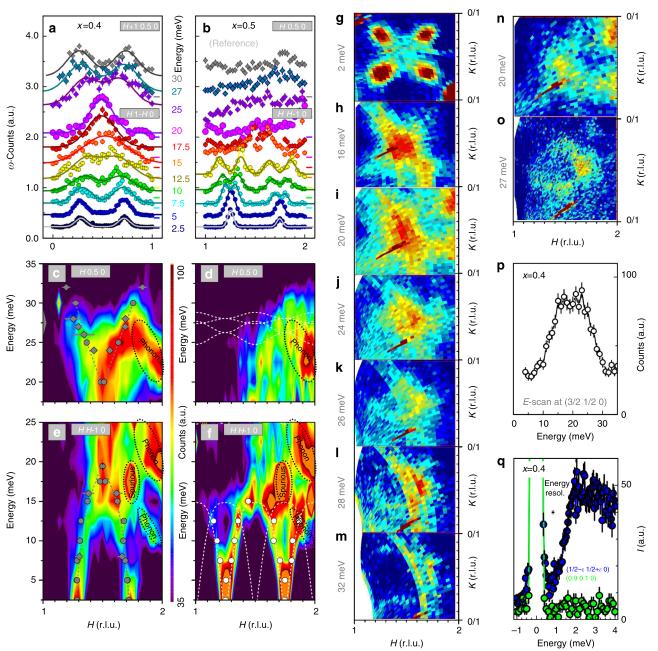}
\end{center}
\caption{\textbf{Neutron spectroscopy study of magnetic excitations}
(a-f) Inelastic neutron scattering intensities for \lscoIV\ and \lscoV\ measured at the 2T spectrometer. (a-b) Neutron scattering intensities (shifted by a value indicated by the horizontal bars); circles: diagonal scans across (1/2~1/2~0) in (a) and (3/2~1/2~0) in (b) as a function of $H$; diamonds: ($H$~0.5~0)-scans plotted as a function of  $H$-1 and $H$ in (a) and (b) respectively; solid lines were derived from symmetric gaussian fits.
(c-f) Intensity maps as a function of $Q$ and energy transfer through $(3/2~1/2~0)$ in $[$$1~0~0$$]$~-~direction (c,d) and $[$$1~1~0$$]$~-~direction (e,f). Maps have been generated from the data above (colours between two data points are interpolated). Dashed lines are guide to the eyes (c,e) and dispersions derived from spin wave calculations (d,f);  black/dashed circles: spurious peaks/phonons.
Error bars for all data included (might be smaller than symbol size). (g-o) Constant-energy maps measured the IN8 spectrometer (energy written to the left of each map). $H$ ($K$) is ranging from 1 to 2 (from 0 to 1) from left to right (from bottom to top) in each map. Instead of interpolation, the whole area around one measured data point is filled with the same colour. Fits of the dispersion (diamonds) which have been derived from these maps are also included in (c,e). In (n,o) the background was subtracted (see supplementary material). (p) Constant-Q scan at ($3/2$~$1/2$~$0$). (q) Constant-Q scans at ($1/2$-$\varepsilon$~$1/2$+$\varepsilon$~$0$) measured at $\sim$10~K at a cold neutron triple-axis spectrometer (blue circles); additionally, the background is shown (green circles). Intensity error bars are statistical error bars calculated by the square root of intensity. Other error bars are standard deviations obtained from Gaussian peak fits.
}
\label{fig2}
\end{figure}

\begin{figure}[ht]

\includegraphics[width=0.8\textwidth]{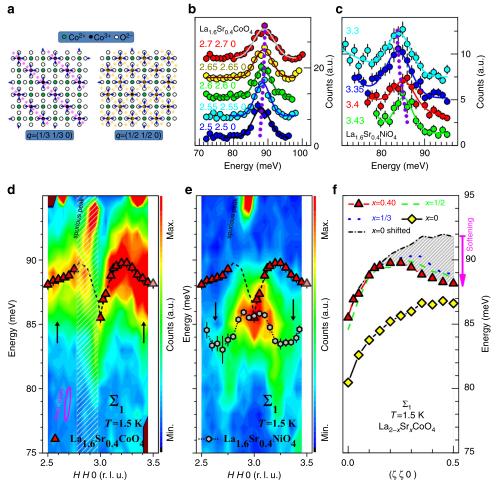}
\begin{center}
\end{center}
\caption{\textbf{Inealstic neutron scattering study of lattice dynamics}\ (a) Polarization patterns for high-frequency $\Sigma_1$ Co-O bond stretching phonon modes. The dashed lines indicate a coupling to charge stripes (left) or to CBCO (right). (b,c) Some representative phonon scans of \lscoIV\ measured at ($3$$-$$q$~$3$$-$$q$~$0$) and for \lsnoIV\ measured at ($3$$+$$q$~$3$$+$$q$~$0$) corresponding to the focusing sides in both compounds. Clearly, the opposite $|q|$-dependence of phonon peak positions can be observed in nickelates and cobaltates. (d) Inelastic neutron scattering intensity as a function of energy and momentum transfer for \lscoIV. The white shaded area is biased by a spurious peak and also measured at the defocusing side. An anomalous phonon softening is clearly observeable at $\zeta$~$=$~$0.5$   rather than at positions indicative for charge stripes (arrows). (e) The corresponding inelastic neutron signal of a 40\% Sr-doped LSNO reference sample with robust diagonal CSO; grey hexagons indicate the fitted phonon dispersion. Unlike in \lscoIV, a phonon softening is observeable around propagation vectors connected to CSO and not for $\zeta$~$\rightarrow$~$0.5$ where even an upturn is observeable.  (f) Comparison of the high-frequency $\Sigma_1$ phonon dispersions of \lsco\ for different $x$. The black dash-dotted line indicates the dispersion of \lco\ shifted to higher energies. Intensity error bars are statistical error bars calculated by the square root of intensity. Other error bars are standard deviations obtained from Gaussian peak fits.}
\label{fig3}
\end{figure}

\begin{figure}[ht]

\begin{center}
\includegraphics[width=0.8\textwidth]{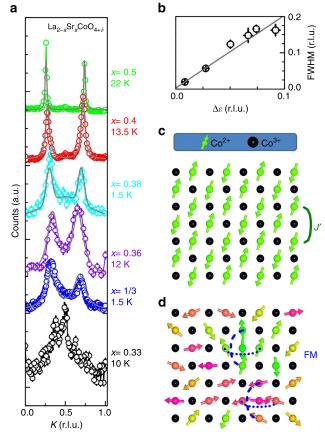}
\end{center}
\caption{\textbf{The frustration scenario in \lsco}\ (a) Neutron scattering intensities for ($H \pm H 0$)-scans of \lsco\ for different $x$ showing a continuous peak broadening below half-doping. (b) The peak width for different $x$ as a function of incommensurability $\Delta\varepsilon$~$=$~$\varepsilon-1/4$. (c) The AFM structure in an ideal CBCO state. (d) Schematic drawing of our frustration scenario. The introduction of an additional electron (\coii-ion) into the AFM matrix of the ideal CBCO half-doped compound induces strong frustration due to a strong nn-exchange coupling $J$ compared to the existing exchange couplings $J'$ within the undistorted AFM matrix. Thus, effectively, one additional ferromagnetic exchange interaction is introduced into each AFM sublattice of the ideal CBCO spin structure (blue dotted/dashed line). In order to release frustration the formation of a chiral or non-collinear magnetic structure appears likely (the change of the spin-directions is emphasized by the colour change). Intensity error bars are statistical error bars calculated by the square root of intensity. Other error bars are standard deviations obtained from Lorentzian peak fits.}
\label{fig1}
\end{figure}


\begin{thebibliography}{10}
\expandafter\ifx\csname url\endcsname\relax
  \def\url#1{\texttt{#1}}\fi
\expandafter\ifx\csname urlprefix\endcsname\relax\def\urlprefix{URL }\fi
\providecommand{\bibinfo}[2]{#2}
\providecommand{\eprint}[2][]{\url{#2}}

\bibitem{stripesA}
\bibinfo{author}{Zaanen, J.} \& \bibinfo{author}{Gunnarsson, O.}
\newblock \bibinfo{title}{Charged magnetic domain lines and the magnetism of
  high-\text{T$_c$} oxides}.
\newblock \emph{\bibinfo{journal}{Phys. Rev. B}} \textbf{\bibinfo{volume}{40}},
  \bibinfo{pages}{7391} (\bibinfo{year}{1989}).

\bibitem{stripesB}
\bibinfo{author}{Schulz, H.~J.}
\newblock \bibinfo{title}{Domain walls in a doped antiferromagnet}.
\newblock \emph{\bibinfo{journal}{J. Phys. (Paris)}}
  \textbf{\bibinfo{volume}{50}}, \bibinfo{pages}{2833} (\bibinfo{year}{1989}).

\bibitem{stripesC}
\bibinfo{author}{Machida, K.}
\newblock \bibinfo{title}{Magnetism in \text{La$_2$CuO$_4$} based compounds}.
\newblock \emph{\bibinfo{journal}{Physica C}} \textbf{\bibinfo{volume}{158}},
  \bibinfo{pages}{192} (\bibinfo{year}{1989}).

\bibitem{stripesLNSCO}
\bibinfo{author}{Tranquada, J.~M.}, \bibinfo{author}{Sternlieb, B.~J.},
  \bibinfo{author}{Axe, J.~D.}, \bibinfo{author}{Nakamura, Y.} \&
  \bibinfo{author}{Uchida, S.}
\newblock \bibinfo{title}{Evidence for stripe correlations of spins and holes
  in copper oxide superconductors}.
\newblock \emph{\bibinfo{journal}{Nature}} \textbf{\bibinfo{volume}{375}},
  \bibinfo{pages}{561} (\bibinfo{year}{1995}).

\bibitem{stripesLSNO}
\bibinfo{author}{Tranquada, J.~M.}, \bibinfo{author}{Buttrey, D.~J.},
  \bibinfo{author}{Sachan, V.} \& \bibinfo{author}{Lorenzo, J.~E.}
\newblock \emph{\bibinfo{journal}{Phys. Rev. Lett.}}
  \textbf{\bibinfo{volume}{73}}, \bibinfo{pages}{1003} (\bibinfo{year}{1994}).

\bibitem{stripesVSsc}
\bibinfo{author}{Tranquada, J.~M.} \emph{et~al.}
\newblock \bibinfo{title}{Simultaneous ordering of holes and spins in
  \text{La$_2$NiO$_{4.125}$}}.
\newblock \emph{\bibinfo{journal}{Phys.Rev.Lett.}}
  \textbf{\bibinfo{volume}{78}}, \bibinfo{pages}{338} (\bibinfo{year}{1997}).

\bibitem{cwik}
\bibinfo{author}{Cwik, M.} \emph{et~al.}
\newblock \bibinfo{title}{Magnetic correlations in
  \text{La$_{2-x}$Sr$_x$CoO$_4$} studied by neutron scattering: Possible
  evidence for stripe phases}.
\newblock \emph{\bibinfo{journal}{Phys. Rev. Lett.}}
  \textbf{\bibinfo{volume}{102}}, \bibinfo{pages}{057201}
  (\bibinfo{year}{2009}).

\bibitem{hourglass}
\bibinfo{author}{Boothroyd, A.~T.}, \bibinfo{author}{Babkevich, P.},
  \bibinfo{author}{Prabhakaran, D.} \& \bibinfo{author}{Freeman, P.~G.}
\newblock \bibinfo{title}{An hour-glass magnetic spectrum in an insulating,
  hole-doped antiferromagnet}.
\newblock \emph{\bibinfo{journal}{Nature}} \textbf{\bibinfo{volume}{471}},
  \bibinfo{pages}{341} (\bibinfo{year}{2011}).

\bibitem{hourglassLSCO}
\bibinfo{author}{Tranquada} \emph{et~al.}
\newblock \bibinfo{title}{Quantum magnetic excitations from stripes in copper
  oxide superconductors}.
\newblock \emph{\bibinfo{journal}{Nature}} \textbf{\bibinfo{volume}{429}},
  \bibinfo{pages}{534} (\bibinfo{year}{2004}).

\bibitem{hourglassLSCOb}
\bibinfo{author}{Matsuda, M.} \emph{et~al.}
\newblock \bibinfo{title}{Magnetic dispersion of the diagonal incommensurate
  phase in lightly doped \text{La$_{2-x}$Sr$_x$CuO$_4$}}.
\newblock \emph{\bibinfo{journal}{Phys. Rev. Lett.}}
  \textbf{\bibinfo{volume}{101}}, \bibinfo{pages}{197001}
  (\bibinfo{year}{2008}).

\bibitem{hourglassLSCOc}
\bibinfo{author}{Arai, M.} \emph{et~al.}
\newblock \bibinfo{title}{Incommensurate spin dynamics of underdoped
  superconductor \text{YBa$_2$Cu$_3$O$_{6.7}$}}.
\newblock \emph{\bibinfo{journal}{Phys. Rev. Lett.}}
  \textbf{\bibinfo{volume}{83}}, \bibinfo{pages}{608} (\bibinfo{year}{1999}).

\bibitem{hourglassLSCOd}
\bibinfo{author}{Bourges, P.} \emph{et~al.}
\newblock \bibinfo{title}{The spin excitation spectrum in superconducting
  \text{YBa$_2$Cu$_3$O$_{6.85}$}}.
\newblock \emph{\bibinfo{journal}{Science}} \textbf{\bibinfo{volume}{288}},
  \bibinfo{pages}{1234} (\bibinfo{year}{2000}).

\bibitem{hourglassLSCOe}
\bibinfo{author}{Christensen, N.~B.} \emph{et~al.}
\newblock \bibinfo{title}{Dispersive excitations in the high-temperature
  superconductor \text{La$_{2-x}$Sr$_x$CuO$_4$}}.
\newblock \emph{\bibinfo{journal}{Phys. Rev. Lett.}}
  \textbf{\bibinfo{volume}{93}}, \bibinfo{pages}{147002}
  (\bibinfo{year}{2004}).

\bibitem{hourglassLSCOf}
\bibinfo{author}{Hayden, S.~M.}, \bibinfo{author}{Mook, H.~A.},
  \bibinfo{author}{Dai, P.}, \bibinfo{author}{Perring, T.~G.} \&
  \bibinfo{author}{Dogan, F.}
\newblock \bibinfo{title}{The structure of the high-energy spin excitations in
  a high-transition-temperature superconductor}.
\newblock \emph{\bibinfo{journal}{Nature}} \textbf{\bibinfo{volume}{429}},
  \bibinfo{pages}{531} (\bibinfo{year}{2004}).

\bibitem{hourglassLSCOg}
\bibinfo{author}{Vignolle, B.} \emph{et~al.}
\newblock \bibinfo{title}{Two energy scales in the spin excitations of the
  high-temperature superconductor \text{La$_{2-x}$Sr$_x$CuO$_4$}}.
\newblock \emph{\bibinfo{journal}{Nature Phys.}} \textbf{\bibinfo{volume}{3}},
  \bibinfo{pages}{163} (\bibinfo{year}{2007}).

\bibitem{hourglassLSCOh}
\bibinfo{author}{Xu, G.} \emph{et~al.}
\newblock \bibinfo{title}{Testing the itinerancy of spin dynamics in
  superconducting \text{Bi$_2$Sr$_2$CaCu$_2$O$_{8+\delta}$}}.
\newblock \emph{\bibinfo{journal}{Nature Phys.}} \textbf{\bibinfo{volume}{5}},
  \bibinfo{pages}{642} (\bibinfo{year}{2009}).

\bibitem{hourglassLSCOi}
\bibinfo{author}{Lipscombe, O.~J.}, \bibinfo{author}{Vignolle, B.},
  \bibinfo{author}{Perring, T.~G.}, \bibinfo{author}{Frost, C.~D.} \&
  \bibinfo{author}{Hayden, S.~M.}
\newblock \bibinfo{title}{Emergence of coherent magnetic excitations in the
  high temperature underdoped \text{La$_{2-x}$Sr$_x$CuO$_4$} superconductor at
  low temperatures}.
\newblock \emph{\bibinfo{journal}{Phys. Rev. Lett.}}
  \textbf{\bibinfo{volume}{102}}, \bibinfo{pages}{167002}
  (\bibinfo{year}{2009}).

\bibitem{pseudo}
\bibinfo{author}{Hinkov, V.} \emph{et~al.}
\newblock \bibinfo{title}{Spin dynamics in the pseudogap state of a
  high-temperature superconductor}.
\newblock \emph{\bibinfo{journal}{Nature Physics}}
  \textbf{\bibinfo{volume}{3}}, \bibinfo{pages}{780} (\bibinfo{year}{2007}).

\bibitem{hourglassExplA}
\bibinfo{author}{Eremin, I.}, \bibinfo{author}{Morr, D.~K.},
  \bibinfo{author}{Chubukov, A.~V.}, \bibinfo{author}{Bennemann, K.~H.} \&
  \bibinfo{author}{Norman, M.~R.}
\newblock \bibinfo{title}{Novel neutron resonance mode in
  \text{d$_{x^2-y^2}$}-wave superconductors}.
\newblock \emph{\bibinfo{journal}{Phys. Rev. Lett.}}
  \textbf{\bibinfo{volume}{94}}, \bibinfo{pages}{147001}
  (\bibinfo{year}{2005}).

\bibitem{hourglassExplB}
\bibinfo{author}{Vojta, M.}, \bibinfo{author}{Vojta, T.} \&
  \bibinfo{author}{Kaul, R.~K.}
\newblock \bibinfo{title}{Spin excitations in fluctuating stripe phases of
  doped cuprate superconductors}.
\newblock \emph{\bibinfo{journal}{Phys. Rev. Lett.}}
  \textbf{\bibinfo{volume}{97}}, \bibinfo{pages}{097001}
  (\bibinfo{year}{2006}).

\bibitem{hourglassExplC}
\bibinfo{author}{Seibold, G.} \& \bibinfo{author}{Lorenzana, J.}
\newblock \bibinfo{title}{Magnetic fluctuations of stripes in the high
  temperature cuprate superconductors}.
\newblock \emph{\bibinfo{journal}{Phys. Rev. Lett.}}
  \textbf{\bibinfo{volume}{94}}, \bibinfo{pages}{107006}
  (\bibinfo{year}{2005}).

\bibitem{hourglassExplD}
\bibinfo{author}{Andersen, B.~M.}, \bibinfo{author}{Graser, S.} \&
  \bibinfo{author}{Hirschfeld, P.~J.}
\newblock \bibinfo{title}{Disorder-induced freezing of dynamical spin
  fluctuations in underdoped cuprate superconductors}.
\newblock \emph{\bibinfo{journal}{Phys. Rev. Lett.}}
  \textbf{\bibinfo{volume}{105}}, \bibinfo{pages}{147002}
  (\bibinfo{year}{2010}).

\bibitem{andrade}
\bibinfo{author}{Andrade, E.~C.} \& \bibinfo{author}{Vojta, M.}
\newblock \bibinfo{title}{Disorder, cluster spin glass, and hourglass spectra
  in striped magnetic insulators}.
\newblock \emph{\bibinfo{journal}{Phys. Rev. Lett.}}
  \textbf{\bibinfo{volume}{109}}, \bibinfo{pages}{147201}
  (\bibinfo{year}{2012}).

\bibitem{highdoped}
\bibinfo{author}{Sakiyama, N.}, \bibinfo{author}{Zaliznyak, I.~A.},
  \bibinfo{author}{Lee, S.-H.}, \bibinfo{author}{Mitsui, Y.} \&
  \bibinfo{author}{Yoshizawa, H.}
\newblock \bibinfo{title}{Doping-dependent charge and spin superstructures in
  layered cobalt perovskites}.
\newblock \emph{\bibinfo{journal}{Phys. Rev. B}} \textbf{\bibinfo{volume}{78}},
  \bibinfo{pages}{180406(R)} (\bibinfo{year}{2008}).

\bibitem{mcphas}
\bibinfo{author}{Rotter, M.}, \bibinfo{author}{Le, D.~M.},
  \bibinfo{author}{Boothroyd, A.~T.} \& \bibinfo{author}{Blanco, J.~A.}
\newblock \bibinfo{title}{Dynamical matrix diagonalization for the calculation
  of dispersive excitations}.
\newblock \emph{\bibinfo{journal}{J. Phys. Cond. Mat.}}
  \textbf{\bibinfo{volume}{24}}, \bibinfo{pages}{213201}
  (\bibinfo{year}{2012}).

\bibitem{half}
\bibinfo{author}{Helme, L.~M.} \emph{et~al.}
\newblock \bibinfo{title}{Magnetic order and dynamics of the charge-ordered
  antiferromagnet \text{La$_{1.5}$Sr$_{0.5}$CoO$_4$}}.
\newblock \emph{\bibinfo{journal}{Phys. Rev. B}} \textbf{\bibinfo{volume}{80}},
  \bibinfo{pages}{134414} (\bibinfo{year}{2009}).

\bibitem{MagLSCOisotropic}
\bibinfo{author}{Matsuda, M.}, \bibinfo{author}{Granroth, G.~E.},
  \bibinfo{author}{Fujita, M.}, \bibinfo{author}{Yamada, K.} \&
  \bibinfo{author}{Tranquada, J.~M.}
\newblock \bibinfo{title}{Energy-dependent crossover from anisotropic to
  isotropic magnetic dispersion in lightly doped la1.96sr0.04cuo4}.
\newblock \emph{\bibinfo{journal}{Phys. Rev. B}} \textbf{\bibinfo{volume}{87}},
  \bibinfo{pages}{054508} (\bibinfo{year}{2013}).

\bibitem{LSCOgap}
\bibinfo{author}{R{\o}mer, A.~T.} \emph{et~al.}
\newblock \bibinfo{title}{Glassy low-energy spin fluctuations and anisotropy
  gap in la1.88sr0.12cuo4}.
\newblock \emph{\bibinfo{journal}{Phys. Rev. B}} \textbf{\bibinfo{volume}{87}},
  \bibinfo{pages}{144513} (\bibinfo{year}{2013}).

\bibitem{LSCOphonon}
\bibinfo{author}{Reznik, D.} \emph{et~al.}
\newblock \bibinfo{title}{Electron-phonon coupling reflecting dynamic charge
  inhomogeneity in copper oxide superconductors}.
\newblock \emph{\bibinfo{journal}{Nature}} \textbf{\bibinfo{volume}{440}},
  \bibinfo{pages}{1170} (\bibinfo{year}{2006}).

\bibitem{dissACK}
\bibinfo{author}{Komarek, A.~C.}
\newblock \emph{\bibinfo{journal}{PhD thesis, Universit\"{a}t zu K\"{o}ln}}
  (\bibinfo{year}{2009}).

\bibitem{zaanen}
\bibinfo{author}{Zaanen, J.}
\newblock \bibinfo{title}{High-temperature superconductivity: The secret of the
  hourglass}.
\newblock \emph{\bibinfo{journal}{Nature}} \textbf{\bibinfo{volume}{471}},
  \bibinfo{pages}{314} (\bibinfo{year}{2011}).

\bibitem{LSCO}
\bibinfo{author}{Sushkov, O.~P.} \& \bibinfo{author}{Kotov, V.~N.}
\newblock \bibinfo{title}{Theory of incommensurate magnetic correlations across
  the insulator-superconductor transition of underdoped
  \text{La$_{2-x}$Sr$_x$CuO$_4$}}.
\newblock \emph{\bibinfo{journal}{Phys. Rev. Lett.}}
  \textbf{\bibinfo{volume}{94}}, \bibinfo{pages}{097005}
  (\bibinfo{year}{2005}).

\bibitem{LSCOb}
\bibinfo{author}{L\"{u}scher, A.}, \bibinfo{author}{Milstein, A.~I.} \&
  \bibinfo{author}{Sushkov, O.~P.}
\newblock \bibinfo{title}{Structure of the spin-glass state of
  \text{La$_{2-x}$Sr$_x$CuO$_4$}: The spiral theory}.
\newblock \emph{\bibinfo{journal}{Phys. Rev. Lett.}}
  \textbf{\bibinfo{volume}{98}}, \bibinfo{pages}{037001}
  (\bibinfo{year}{2007}).

\bibitem{hasselmann}
\bibinfo{author}{Hasselmann, N.}, \bibinfo{author}{\text{Castro~Neto}, A.~H.}
  \& \bibinfo{author}{\text{Morais~Smith}, C.}
\newblock \bibinfo{title}{Spin-glass phase of cuprates}.
\newblock \emph{\bibinfo{journal}{Phys. Rev. B}} \textbf{\bibinfo{volume}{69}},
  \bibinfo{pages}{014424} (\bibinfo{year}{2004}).

\bibitem{ourLSCO}
\bibinfo{author}{Wu, H.-H.} \emph{et~al.}
\newblock \bibinfo{title}{Charge stripe order near the surface of 12-percent
  doped \text{La$_{2-x}$Sr$_x$CuO$_4$}}.
\newblock \emph{\bibinfo{journal}{Nature Commun.}}
  \textbf{\bibinfo{volume}{3}}, \bibinfo{pages}{1023} (\bibinfo{year}{2012}).

\bibitem{nacoo2}
\bibinfo{author}{Takada, K.} \emph{et~al.}
\newblock \bibinfo{title}{Superconductivity in two-dimensional \text{CoO$_2$}
  layers}.
\newblock \emph{\bibinfo{journal}{Nature}} \textbf{\bibinfo{volume}{422}},
  \bibinfo{pages}{53} (\bibinfo{year}{2003}).

\bibitem{growthLSCoO}
\bibinfo{author}{Matsuura, T.}, \bibinfo{author}{Mizusaki, J.},
  \bibinfo{author}{Yamauchi, S.} \& \bibinfo{author}{Fueki, K.}
\newblock \bibinfo{title}{Single crystal growth of
  \text{La$_{2-x}$Sr$_x$CoO$_4$} ($x$=0.0, 0.5, 1.0 and 1.5)}.
\newblock \emph{\bibinfo{journal}{Jpn. J. Appl. Phys.}}
  \textbf{\bibinfo{volume}{23}}, \bibinfo{pages}{1143} (\bibinfo{year}{1984}).

\bibitem{LSNOcryst}
\bibinfo{author}{Jang, W.-J.} \& \bibinfo{author}{Takei, H.}
\newblock \bibinfo{title}{Growth, structure and properties of
  \text{La$_{2-x}$Sr$_x$NiO$_4$} ($x$= 0 to 0.3) single crystals}.
\newblock \emph{\bibinfo{journal}{Jpn. J. Appl. Phys.}}
  \textbf{\bibinfo{volume}{30}}, \bibinfo{pages}{251} (\bibinfo{year}{1991}).

\end{thebibliography}
\end{document}